\documentclass[12pt, preprint]{aastex}

\usepackage{color}
\usepackage{hyperref}
\newcommand{\niceurl}[1]{\href{#1}{\textsl{#1}}}
\newcommand{\viewerurl}{\niceurl{http://legacysurvey.org/viewer}}

\begin{document}

\title{The X-shaped Bulge of the Milky Way revealed by WISE}
\author{%
Melissa Ness\altaffilmark{1,2} \& 
Dustin Lang\altaffilmark{3,4}
}
\shortauthors{Ness, Lang}
\altaffiltext{1}{%
  Max-Planck-Institut f\"ur Astronomie,
  K\"onigstuhl 17, D-69117 Heidelberg, Germany
}
\altaffiltext{2}{%
  To whom correspondence should be addressed:
  ness@mpia-hd.mpg.de
}
\altaffiltext{3}{%
  Dunlap Institute and Department of Astronomy \& Astrophysics,
  University of Toronto,
  50 Saint George Street, Toronto, ON, M5S 3H4, Canada
}
\altaffiltext{4}{%
  Department of Physics \& Astronomy,
  University of Waterloo,
  200 University Avenue West, Waterloo, ON, N2L 3G1, Canada
}
\begin{abstract}%
The Milky Way bulge has a boxy/peanut morphology and an X-shaped structure. This X-shape has been revealed by the `split in the red clump' from 
star counts along the line of sight toward the bulge, measured from photometric surveys. This boxy, X-shaped bulge morphology is not unique to the Milky Way and such bulges are observed in other barred spiral galaxies.  N-body simulations show that boxy and X-shaped bulges are formed from the disk via dynamical instabilities.  It has also been proposed that the Milky Way bulge is not X-shaped, but rather, the apparent split in the red clump stars is a consequence of different stellar populations, in an old classical spheroidal bulge. We present a WISE image of the Milky Way bulge, produced by downsampling the publicly available ``unWISE'' coadds. The WISE image of the Milky Way bulge shows that the X-shaped nature of the Milky Way bulge is self-evident and irrefutable. The X-shape morphology of the bulge in itself and the fraction of bulge stars that comprise orbits within this structure has important implications for the formation history of the Milky Way, and, given the ubiquity of boxy X-shaped bulges, spiral galaxies in general. 
\end{abstract}


\section{Introduction}\label{sec:Intro}

The boxy nature of the bulge of the Milky Way was first revealed in
the COBE satellite image \citep{Dwek1995}. Using photometric data from
the VVV survey, \citet{Wegg2013} measured the three dimensional
density of red clump stars in the bulge, earlier revealed to show two
peaks along the line of sight \citep{McWilliam2010, Nataf2010}, and
determined their distribution to be characteristic of a strong
boxy/peanut bulge within a barred galaxy. This observed split in the
red clump is reported to be a property of the more metal rich stars in
the bulge, with [Fe/H] $>$ --0.5 \citep{Ness2012, Uttenthaler2012};
although according to \citet{Nataf2014} this metallicity dependence
may be subject to biases.  \citet{Portail2015a} used the VVV red clump
stellar density to show that the Milky Way's bulge has an off-centered
X-structure and using orbit based characterization of the
X-shape, determined that the fraction of stars in orbits that
contribute to the X-shape is 40--45\% of the mass of the bulge
\citep{Portail2015b}. The X-shape in the Milky Way bulge is similar to
that seen in the unsharp masked images of other barred spiral galaxies
\citep[e.g.][]{Bureau2006}. Such X-shaped structures, which underly
the boxy/peanut, have been shown to form in N-body simulations, via
dynamical instabilities in the disk \citep[e.g.][]{Athanassoula2005,
  Debattista2006, Inma2006}. This shape is a consequence of the
x$_{1}$v$_{1}$ \citep{P1984, Athanassoula1992} and other orbit
families \citep[e.g.][]{Portail2015b}.

\citet{Lee2015} have questioned the existence of the X-shaped nature of the bulge, instead proposing the split in the red clump stars to be a consequence of different stellar populations in a classical bulge. Conversely,  \citet{Gonzalez2015} summarise the set of observational properties of the bulge which explain the link between the double or split red clump and the X-shape.

 We present, for the first time, the WISE image of the Milky Way \citep{Lang2014a} which clearly demonstrates the Milky Way bulge is irrefutably morphologically, X-shaped. This follows expectations from the observational evidence,  from dynamical models of boxy/peanut bulges like the Milky Way, and from observations of other barred galaxies that reveal such an X-shaped profile is not uncommon \citep{L2014}. 

\section{The unWISE image of the Milky Way bulge}

The Wide-Field Infrared Survey Explorer \citep[WISE;][]{W2010} is a full sky photometric survey using four bands in the mid-infrared at 3.4 $\micron$, 4.6 $\micron$, 12 $\micron$ and 22 $\micron$ (W1--W4). The original WISE data release was based on a co-adding approach that is optimal for detecting isolated point sources but which effectively blurs the images. \citet{Lang2014a} implemented an alternative co-adding methodology that does not degrade the resolution of the imaging.  These ``unWISE'' coadds have been publicly released.\footnote{Available at \niceurl{http://unwise.me}. Browseable at \viewerurl.}

\begin{figure}[h!]
\centering
        \includegraphics[width=\textwidth]{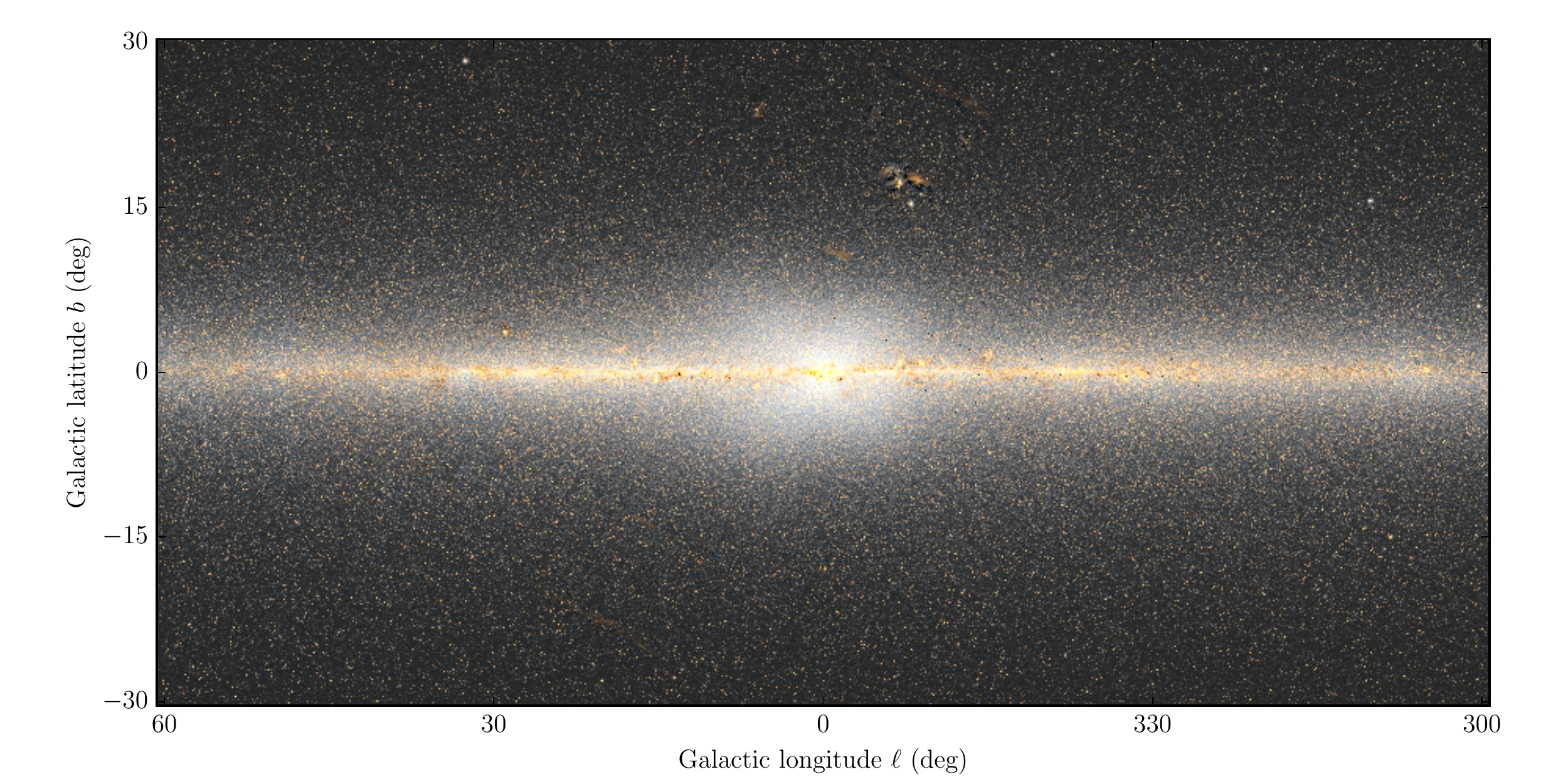}
\caption{WISE image for W1 and W2 in Galactic coordinates.  An arcsinh
  stretch is used to allow the full dynamic range to be shown.}
\label{fig:xbulge}
\end{figure}

Figure \ref{fig:xbulge} presents the bulge region of the Milky Way,  resampled to galactic coordinates from the unWISE images, across $(|\ell|,|b|)$ $<$ (60, 30) in WISE bands W1 and W2.  This presents the overall Milky Way structures in exquisite detail.
No additional unsharp masking or equivalent techniques have been used to enhance these data. The bulge in the central region of this Figure shows a clear X-shaped morphology. Note that the arms of the X-shape are asymmetrical around the minor axis and appear larger at left than at right. This is a real projection effect and reflects that the bulge is oriented at about 27$^\circ$ degrees with respect to the line of sight \citep{Wegg2013}, with the nearest side at positive longitudes. The X-shape is also visible, though with more artifacts, in the official AllWISE data release imaging.\footnote{Explanatory
  Supplement to the AllWISE Data Release Products, 
  \niceurl{http://wise2.ipac.caltech.edu/docs/release/allwise/expsup/}}
The W3 and W4 bands largely trace dust rather than stellar light and therefore
do not reveal the X-shaped profile.

\section{The contrast enhanced unWISE image of the bulge}

\begin{figure}[h!]
\centering
        \includegraphics[width=\textwidth]{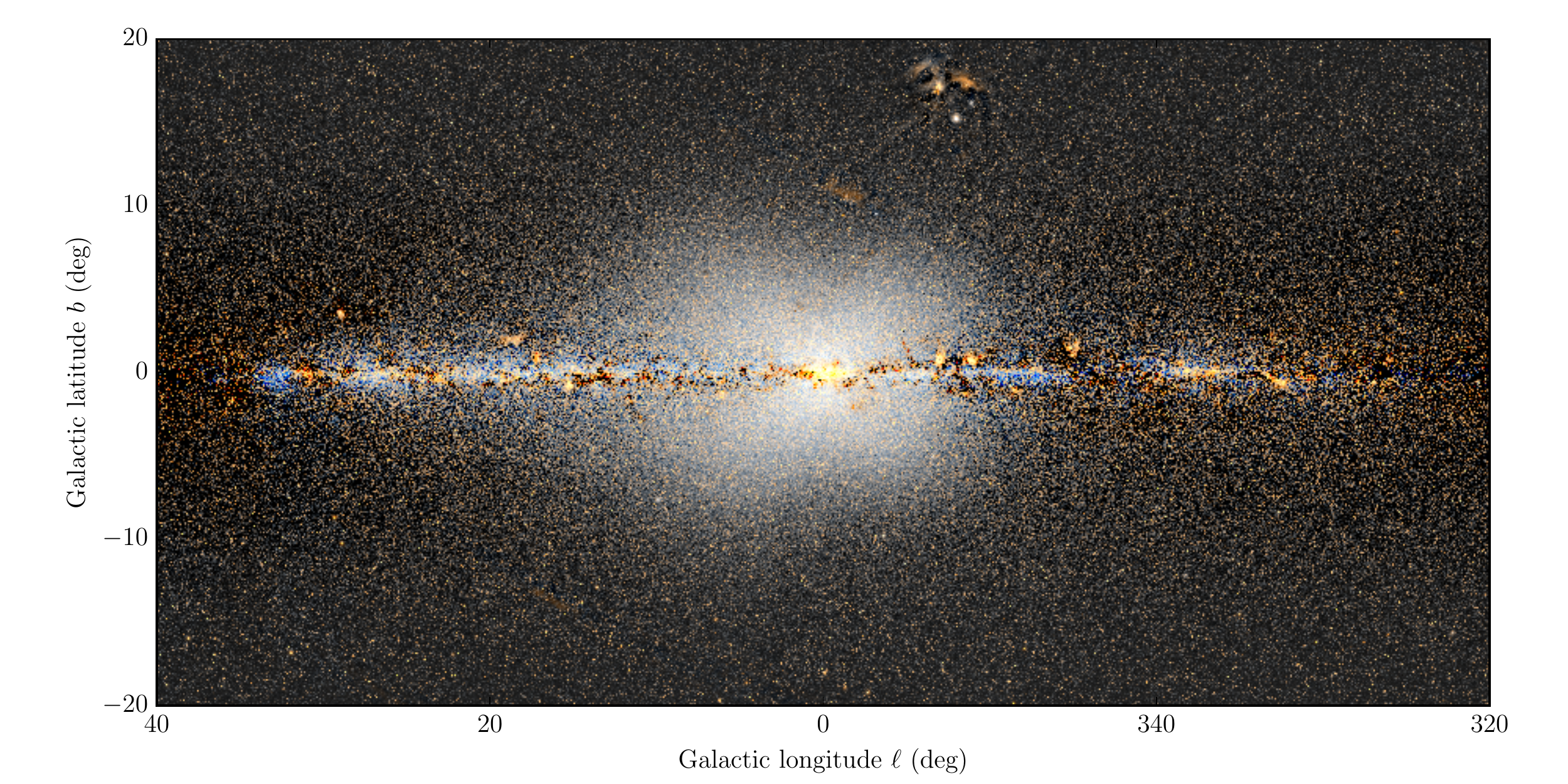}
\caption{The same as Figure 1, but zoomed in and with the median of each row of the image subtracted to provide a better contrast which reveals the X-shape morphology in better detail.}
\label{fig:filt}
\end{figure}

Figure \ref{fig:filt} presents a contrast enhanced and zoomed in version of Figure \ref{fig:xbulge}. This better reveals the X-shape light profile of the bulge and its extent across $(\ell,b)$ in the WISE image. This Figure was produced with a median subtraction across each row to suppress the contribution from the disk. The arms in the image extend to longitudes of $|\ell|$ $\approx$ 10$^\circ$ and latitudes of $|b|$ $\lesssim$ 10$^\circ$; (although note again the arms on the near side are larger than those of the far side due to projection). This extent on the sky corresponds to a length of about 2.4 kpc for each arm, for a distance of the bulge of 8.3 kpc from the Sun and a bar angle of 27$^\circ$.

\section{The residuals of the unWISE image of the bulge}

\begin{figure}[h!]
\centering
\includegraphics[width=0.3\textwidth]{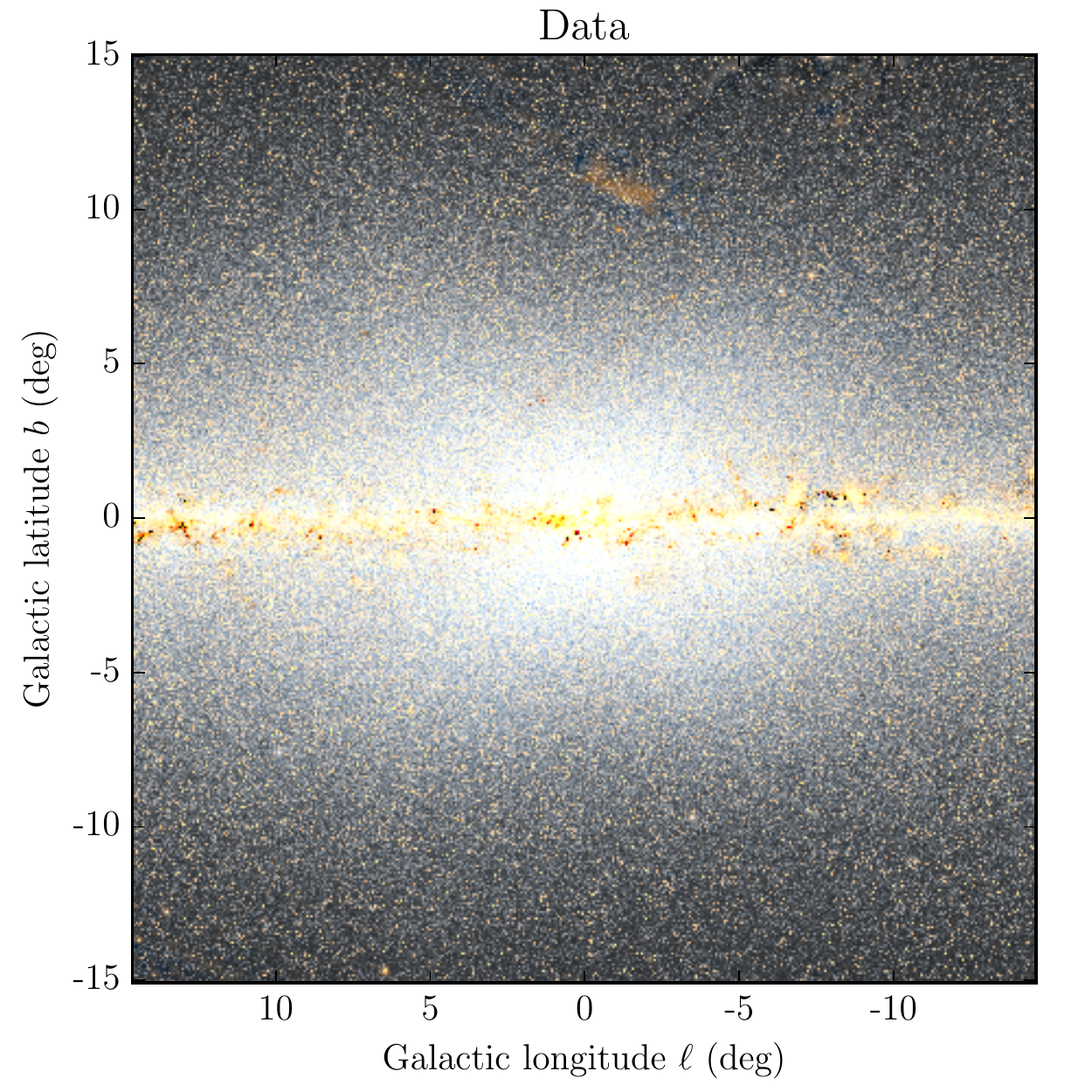}
\includegraphics[width=0.3\textwidth]{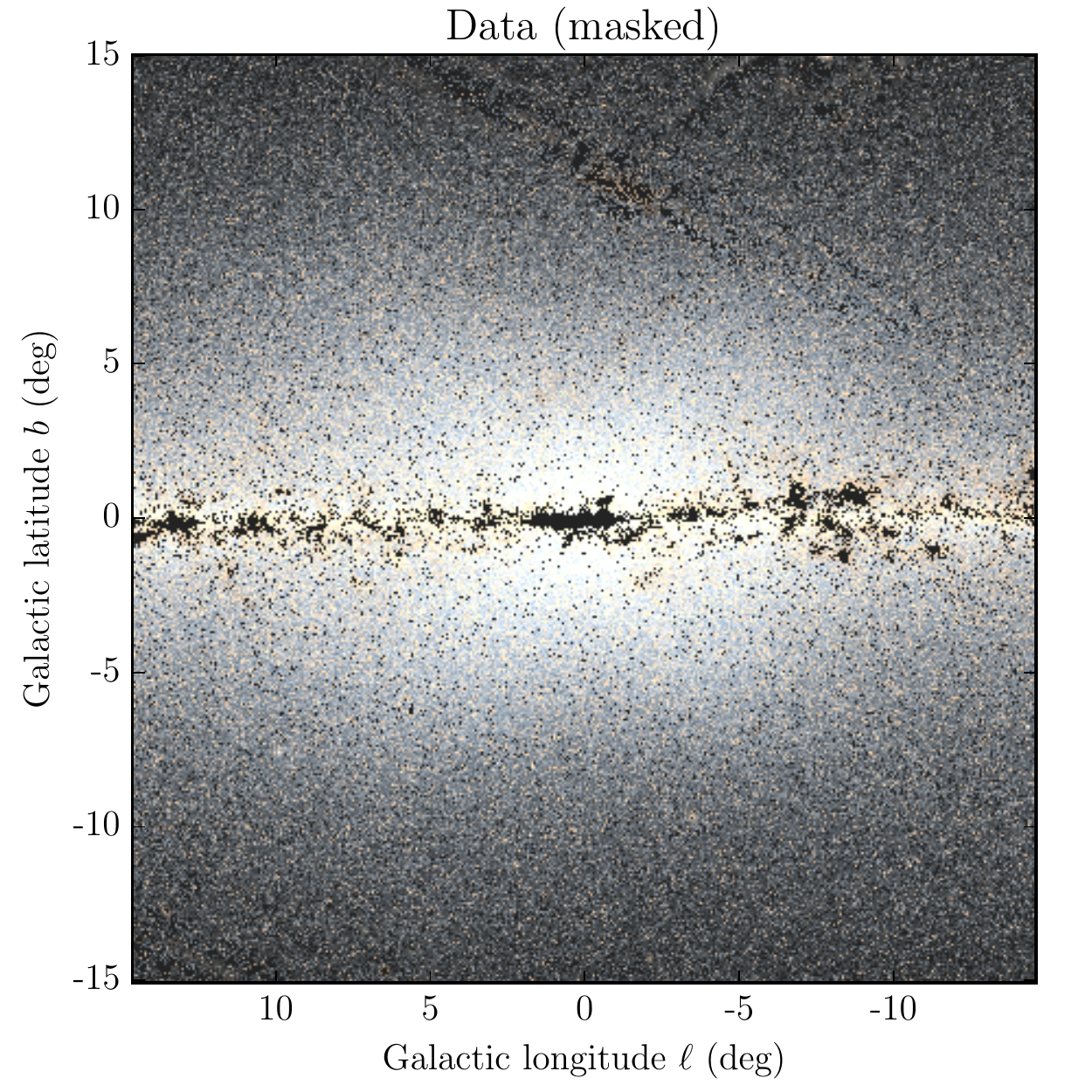}
\includegraphics[width=0.3\textwidth]{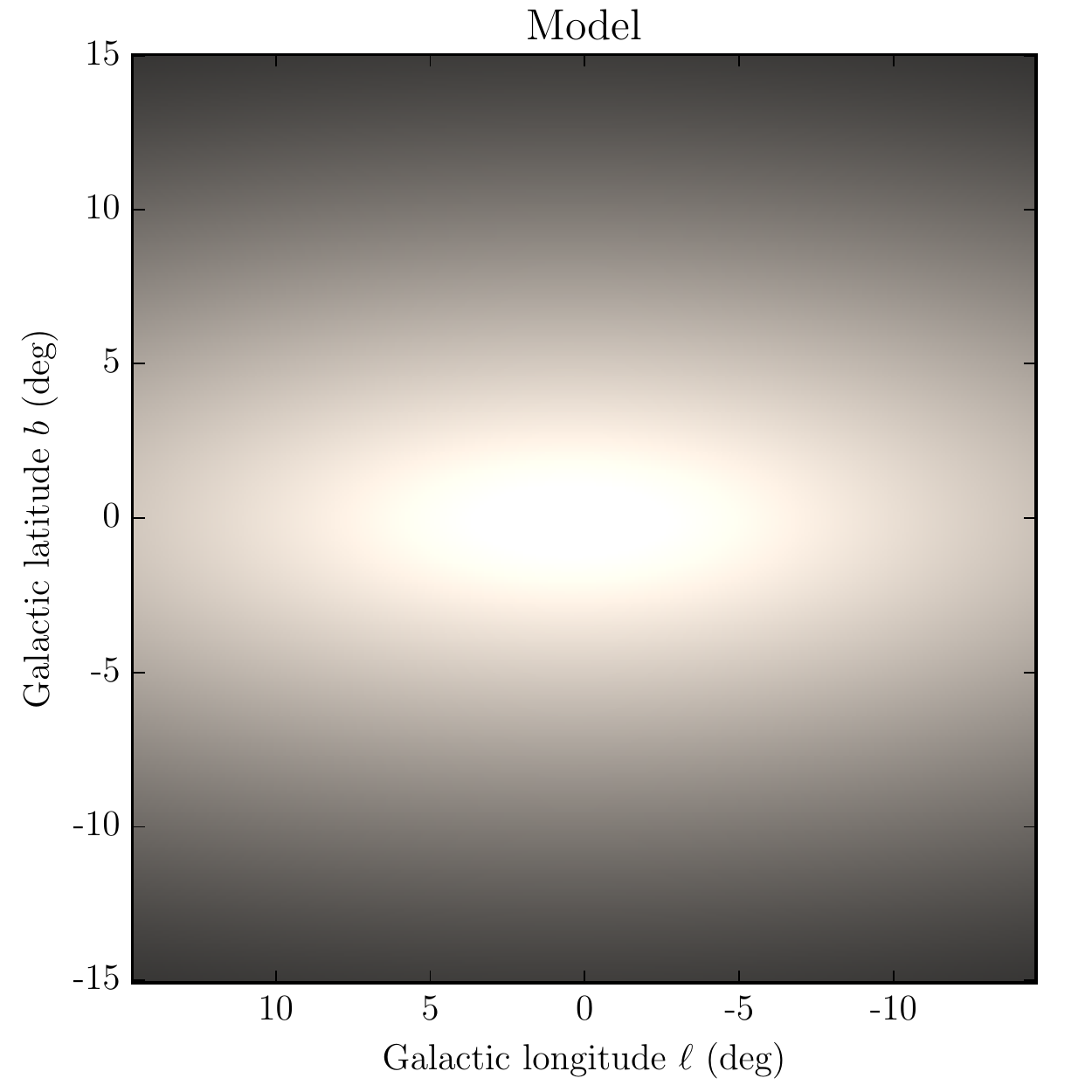}
\\
\includegraphics[width=0.3\textwidth]{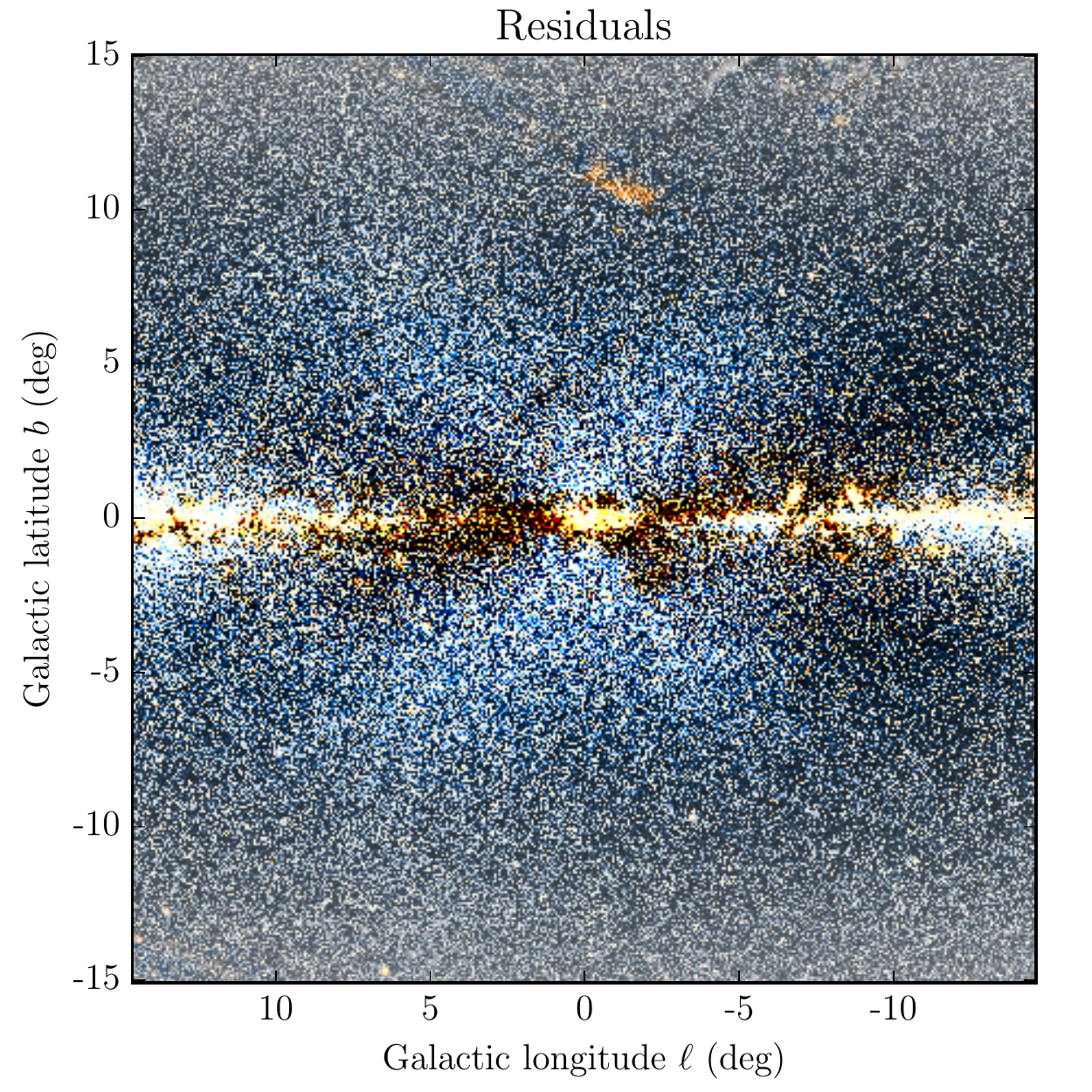}
\includegraphics[width=0.3\textwidth]{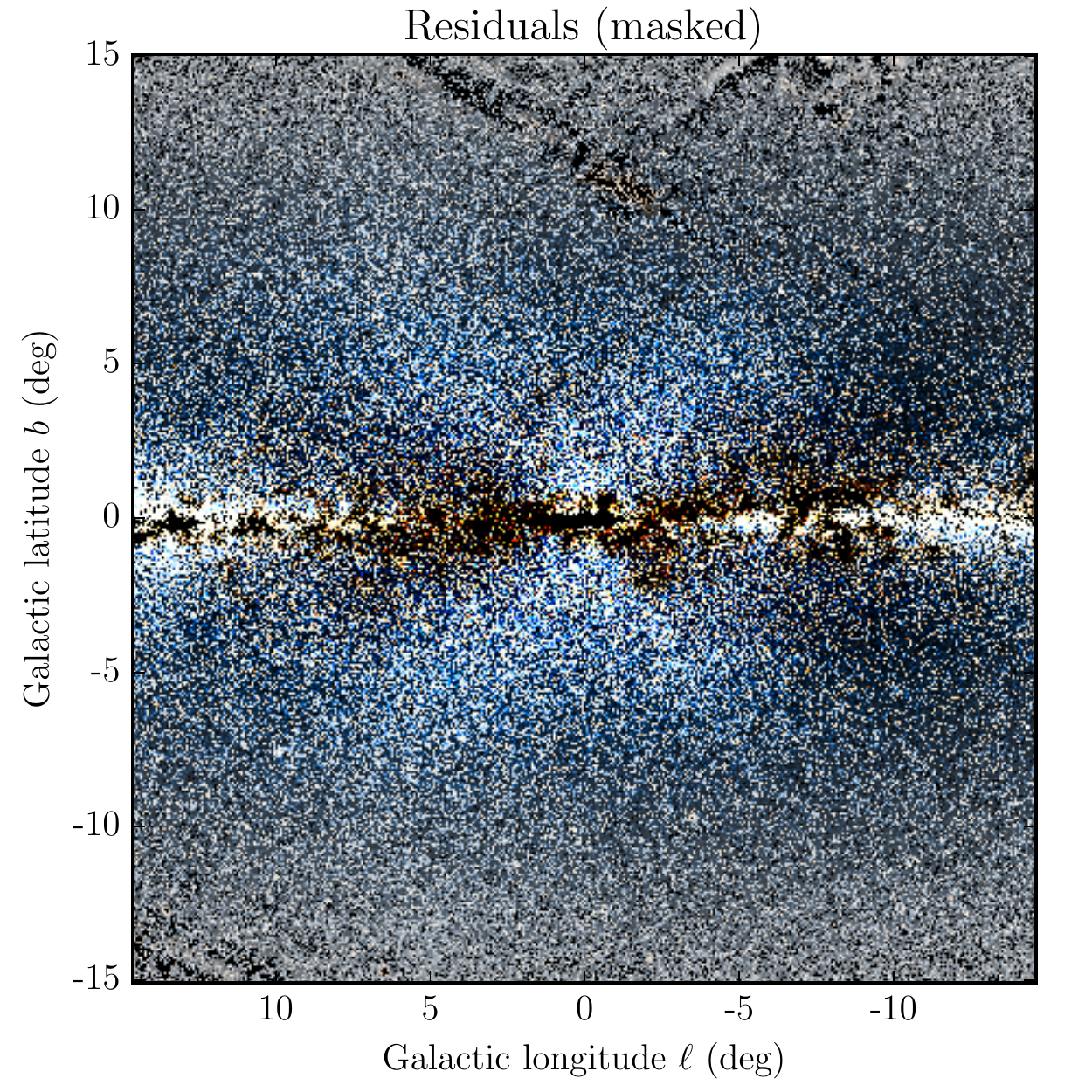}
\includegraphics[width=0.3\textwidth]{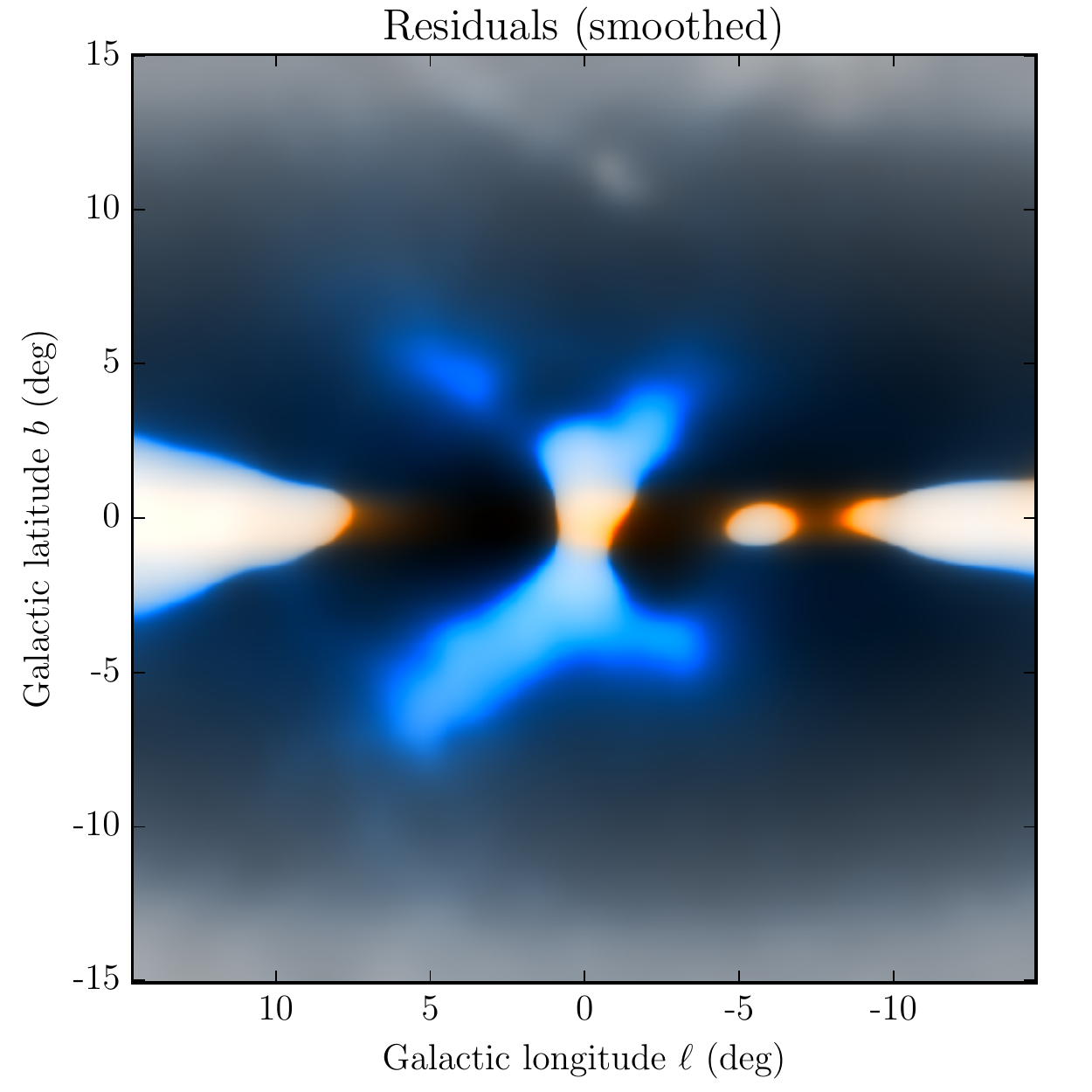}
\caption{%
  The WISE W1 and W2 image fit by a simple exponential disk
  model, making the X structure more apparent.
  Top-left: Data.  Top-middle: Data, masking out the top and bottom 5\%
  of pixels based on $\textrm{W1} - \textrm{W2}$ color, as well as pixels with negative flux.  The diagonal structure
  at the top of the image is due to scattered light from the Moon in the unWISE coadds.
  Top-right: Exponential disk model fit.
  Bottom-left: Residuals (data minus model).  Bottom-middle: Masked residuals.
  Bottom-right: 50-pixel ($\sim 1.7^{\circ}$) median filter of masked residuals (median of unmasked
  pixels).
  }
\label{fig:modfit}
\end{figure}

Finally, we fit and subtract a simple exponential disk model to make the bulge structure more clear.  We zoom in to the central region of the galaxy, compute the $\textrm{W1} - \textrm{W2}$ color of each pixel, and mask the top and bottom 5\% in order to suppress the influence of the most dusty regions on the fit.  We then fit a simple exponential disk model, where the ellipse shape parameteres are shared between the W1 and W2 bands, and each pixel is given equal weight.
The model fit parameters we get are a half-light radius on the major axis of about 1.9 kpc and an axis ratio of 0.38, yielding a vertical half-light radius of about 720 pc.
Figure \ref{fig:modfit} shows these results.
The projection effect is again clear in these residual maps, which reflects that the bulge is orientated at 27$^\circ$ with respect to the line of sight, with the nearest side at  positive longitudes. %

\section{Conclusion}

Using the publicly available ``unWISE'' coadds, we have presented new images of the bulge of the Milky Way in the W1 and W2 bands.  These directly reveal its X-shaped nature in the integrated light, even without any special image processing or enhancement. Our contrast enhancement and residual maps further highlight the extent of the X-shape that underlies the boxy structure. Critical to understanding the bulge is a further and detailed characterisation of the stars that are in the arms of the X-shape, the spatial extent of which is clearly demonstrated in our Figures. The spatial mapping we have presented in galactic coordinates will therefore provide a useful guide for current and future spectroscopic surveys such as APOGEE \citep{Majewski2015}, 4-MOST \citep{4most} and bulge programs associated with the GALAH survey \citep{deSilva2015}. These data can be used to guide stellar target selection, where examining the spectroscopic ages \citep[e.g.][]{Martig2016, Ness2016} and metallicities of stars in the arms of the X-shape as a function of $(\ell,b)$ will be necessary to understand the formation of the bulge and constrain the formation processes relevant in the Milky Way.

\section{Acknowledgements} 
We thank Hans-Walter Rix (MPIA) for helpful comments on this work. 
The research has received funding from the European Research Council under the European Union's Seventh Framework Programme (FP 7) ERC Grant Agreement n.~[321035].
The Dunlap Institute is funded through an endowment established by the David Dunlap family and the University of Toronto.

\bibliography{Xbulge_bib.bib}

\end{document}